\documentclass[%
 aip,
 jmp,%
 amsmath,amssymb,
 reprint,%
]{revtex4-1}

\usepackage{graphicx}
\usepackage{dcolumn}
\usepackage{bm}

\begin{document}

\preprint{AIP/123-QED}

\title[Hydrodynamic Approach to the Free Electron Laser Instability]{Hydrodynamic Approach to the Free Electron Laser Instability\footnote{Error!}}

\author{Stephan I. Tzenov}

\affiliation{Department of Physics, Lancaster University, Lancaster LA1 4YB, United Kingdom}

\altaffiliation{The Cockcroft Institute, Keckwick Lane, Daresbury, WA4 4AD, United Kingdom.}


\email{s.tzenov@lancs.ac.uk}


\author{Kiril B. Marinov}

\affiliation{ASTeC, STFC Daresbury Laboratory, Keckwick Lane, Daresbury, WA4 4AD, United Kingdom.}

\altaffiliation{The Cockcroft Institute, Keckwick Lane, Daresbury, WA4 4AD, United Kingdom.}


\date{\today}

\begin{abstract}
Starting from the Vlasov-Maxwell system, an exact relativistic hydrodynamic closure for a special type water bag distributions satisfying the Vlasov equation has been derived. It has been shown that the hydrodynamic equations are fully equivalent to the original Vlasov-Maxwell equations. The linear stability of the exact hydrodynamic closure has been studied as well. It has been shown that all basic features of the small signal gain can be derived from the fluid dynamic description developed here. Satisfactory agreement with previously reported results has been also found.
\end{abstract}

\maketitle

%

\section{\label{sec:intro}Introduction}

Conventional lasers are ubiquitous sources of coherent electromagnetic radiation over a wide range of the spectrum, from the infrared (around 10 mm) down to the ultraviolet (around 200 nm). However, in the far-infrared, the so-called terahertz part of the spectrum (in the range from 30 mm to 1 mm), or at X-ray wavelengths (less than 10 nm), traditional lasers are not easily achievable. There is therefore ever growing interest in alternate sources of intense, ultra fast, coherent radiation
covering these two portions of the electromagnetic spectrum.

The free electron laser \cite{Saldin,Freund} (FEL) is a relatively new kind of laser, in which the electrons are not bound in atomic or molecular systems, neither they are confined by a lattice. The source of these 'free' electrons is an electron beam accelerated to suitable energy in a linac or a synchrotron. Generally speaking, the free electron laser converts part of the kinetic energy of the electrons into coherent electromagnetic radiation. This conversion is facilitated by a static magnetic field produced by a magnetic device called an undulator \cite{Clarke}. Free electron lasers can produce continuous and widely tunable coherent radiation in any part of the electromagnetic spectrum. In addition, very short pulses can be emitted at any wavelength with the laser intensity being very high.

A number of theoretical models have been developed over the years in order to describe the free electron laser. Although the first analyses involved techniques from quantum mechanics \cite{Madey} and quantum electrodynamics \cite{Colson,Becker}, classical methods have proved to provide a clear and accurate picture of the underlying processes \cite{Bambini,Kroll,Kwan,Ride}. The self consistent Maxwell-Boltzmann equations have been further transformed into quasi-Bloch equations in order to stress on the laser physics perspective of the problem \cite{Hopf}. The plasma physics nature of the electron dynamics moving in the undulator field has been pursued in the kinetic description based on the coupled Vlasov-Maxwell equations; an approach that has been developed by Davidson and coworkers \cite{Davidson,DavidsonUhm}. Many features of the dynamics of free electron lasers can be understood using a simple pendulum model. It comes about because the electromagnetic wave of the radiated field and the magnetic field of the undulator act in tandem on the electron to produce a sinusoidal potential similar to that of a pendulum \cite{Piovella,Bonifacio,Bonifacio1}. This model is to a large extent similar to the so-called single-wave model, widely used in conventional plasma physics \cite{MarTze}.

A relativistic cold-fluid dynamics model of an electron beam with uniform cross section propagating axially through a constant-amplitude helical wiggler magnetic field has been developed as well \cite{DavidsonSen,Sen}. The present paper can be considered as a further extension in this direction.

Starting from very basic principles of propagation of a relativistic electron in the static magnetic field of an undulator, and assuming that its dynamics depends on the longitudinal coordinate and the time only, we arrive at a coupled Vlasov-Maxwell system of equations in one spatial dimension and time. Taking into account an exact solution of the Vlasov equation in the form of a relativistic water bag distribution, we preform in Section \ref{sec:hydro} a reduction of the Vlasov-Maxwell system to an exact closure of relativistic fluid dynamic equations coupled with the wave equations for the radiation field. An interesting feature of the hydrodynamic system thus derived is the fact that the ponderomotive potential together with the pressure law enter the picture in the form of an effective enthalpy. In Section \ref{sec:linear}, we present the linear stability analysis in the derived fluid dynamic framework, which is fully equivalent to the kinetic description in terms of the self-consistent Vlasov-Maxwell equations. The numerical solution of the dispersion equation derived in Section \ref{sec:linear} is presented in Section \ref{sec:simulation} and certain useful properties of the imaginary part of its roots (the so-called small-signal gain) are discussed. Finally, in Section \ref{sec:remarks}, we draw some conclusions.

\section{\label{sec:generalities}Generalities and First Principles}

The present analysis assumes a relativistic electron beam propagating in longitudinal $s$ direction through a helical wiggler magnetic field described by
\begin{equation}
{\bf B}_w = B_0 {\left( {\bf e}_x \cos k_0 s + {\bf e}_y \sin k_0 s \right)}, \label{Helmagfield}
\end{equation}
\noindent where $B_0 = {\rm const}$ is the field amplitude, $\lambda_0 = 2 \pi / k_0$ is the wavelength, and ${\bf e}_x$ and ${\bf e}_y$ are unit Cartesian vectors in the plane perpendicular to the propagation direction. The vector potential associated with the wiggler field (\ref{Helmagfield}) is given by
\begin{equation}
{\bf A}_w = - {\frac {B_0} {k_0}} {\left( {\bf e}_x \cos k_0 s + {\bf e}_y \sin k_0 s \right)}. \label{Vecpot}
\end{equation}
\noindent In what follows we assume that spatial variations are one-dimensional in nature, so that $\partial / \partial x = \partial / \partial y = 0$ and $\partial / \partial s$ is generally nonzero.

The Hamiltonian of an electron moving in the wiggler field, as well as in the self-consistent electromagnetic field can be written as
\begin{equation}
{\cal H} = m \gamma c^2  = c {\sqrt{m^2 c^2 + {\left( {\bf P} + e {\bf A} \right)}^2}} - e \Phi, \label{Hamiltonian}
\end{equation}
\noindent where $m$ is the electron rest mass, $\gamma$ is the relativistic Lorentz factor, $c$ is the velocity of light in vacuo and ${\bf P}$ is the canonical momentum. In addition, $\Phi$ is the scalar self-potential and ${\bf A}$ is the total vector potential ${\bf A} = {\bf A}_w + {\bf A}_r$, where ${\bf A}_r$ describes the radiation self-field. It is convenient to pass to dimensionless momenta and new "time", dimensionless Hamiltonian, and electromagnetic potentials
\begin{equation}
{\bf p} = {\frac {\bf P} {m c}}, \qquad \qquad \tau = c t, \label{Dimomtime}
\end{equation}
\begin{equation}
H = \gamma  = {\sqrt{1 + {\left( {\bf p} + {\bf a} \right)}^2}} - \varphi, \label{Dimham}
\end{equation}
\begin{equation}
{\bf a} = {\frac {e {\bf A}} {m c}}, \qquad \qquad \varphi = {\frac {e \Phi} {m c^2}}. \label{Dimpoten}
\end{equation}
\noindent In order to eliminate the longitudinal component of the vector potential $a_s$ under the square root, we redefine the longitudinal component of the particle momentum by means of a canonical transformation specified by the generating function
\begin{equation}
F_2 {\left( {\bf x}, {\widetilde{\bf p}}; \tau \right)} = x {\widetilde p}_x + y {\widetilde p}_y + s {\widetilde p}_s - \int {\rm d} s a_s {\left( s; \tau \right)} . \label{Canontrans}
\end{equation}
\noindent Dropping the tilde in what follows, we write the new Hamiltonian
\begin{equation}
\gamma  = {\sqrt{1 + p_s^2 + {\left( {\bf p} + {\bf a} \right)}_{\perp}^2}} - \varphi - {\frac {\partial} {\partial \tau}} \int {\rm d} s a_s {\left( s; \tau \right)}. \label{Dimhamiltonian}
\end{equation}
\noindent Here the subscript "${\perp}$" corresponds to the transverse components of the canonical coordinates and fields.

Hamilton's equations of motion can be written as
\begin{equation}
{\frac {{\rm d} {\bf x}_{\perp}} {{\rm d} \tau}} = {\frac {{\bf p}_{\perp} + {\bf a}_{\perp}} {\gamma_c}}, \qquad \qquad {\frac {{\rm d} {\bf p}_{\perp}} {{\rm d} \tau}} = 0, \label{Hamilequmot1}
\end{equation}
\begin{equation}
{\frac {{\rm d} s} {{\rm d} \tau}} = {\frac {p_s} {\gamma_c}}, \qquad {\frac {{\rm d} p_s} {{\rm d} \tau}} = - {\frac {\partial \gamma_c} {\partial s}} + {\cal F}, \label{Hamilequmot2}
\end{equation}
\noindent where $\gamma_c$ denotes the kinetic term (the square root) in the total Hamiltonian (\ref{Dimhamiltonian}). In addition, ${\cal F}$ is the electric force
\begin{equation}
{\cal F} = {\frac {\partial \varphi} {\partial s}} + {\frac {\partial a_s} {\partial \tau}}, \label{Elecforce}
\end{equation}
\noindent acting on the particle. In the present geometry, there are two exact single-particle invariants in the combined external ${\bf A}_w$ and self-fields ${\bf A}_r$ and $\Phi$ configuration. These are the canonical momenta ${\bf p}_{\perp}$, transverse to the beam propagation direction.

The nonlinear Vlasov equation for the distribution function $f {\left( {\bf x}, {\bf p}; \tau \right)}$ can be written as
\begin{equation}
{\frac {\partial f} {\partial \tau}} + {\frac {{\bf p}_{\perp} + {\bf a}_{\perp}} {\gamma_c}} \cdot {\frac {\partial f} {\partial {\bf x}_{\perp}}} + {\frac {p_s} {\gamma_c}} {\frac {\partial f} {\partial s}} + {\left( {\cal F} - {\frac {\partial \gamma_c} {\partial s}} \right)} {\frac {\partial f} {\partial p_s}} = 0. \label{Vlasovequ}
\end{equation}
\noindent It possesses an exact solution of the form
\begin{equation}
f {\left( {\bf x}, {\bf p}; \tau \right)} = \delta {\left( p_x \right)} \delta {\left( p_y \right)} F {\left( s, p_s; \tau \right)}, \label{Solvlasov}
\end{equation}
\noindent where $p_x$ and $p_y$ are the exact invariants, defined by the last two of Eqs. (\ref{Hamilequmot1}). Expression (\ref{Solvlasov}) implies that the transverse motion of the electrons is "cold"  since the transverse beam emittance has been neglected. The evolution of the yet unknown function $F {\left( s, p_s; \tau \right)}$ of the longitudinal canonical coordinates and time is governed by the one-dimensional Vlasov equation
\begin{equation}
{\frac {\partial F} {\partial \tau}} + {\frac {p_s} {\gamma_c}} {\frac {\partial F} {\partial s}} + {\left( {\cal F} - {\frac {\partial \gamma_c} {\partial s}} \right)} {\frac {\partial F} {\partial p_s}} = 0, \label{Onedimvlasov}
\end{equation}
\noindent where $\gamma_c$ defined as
\begin{equation}
\gamma_c {\left( s, p_s; \tau \right)} = {\sqrt{1 + p_s^2 + a^2 {\left( s; \tau \right)}}}, \label{Gammac}
\end{equation}
\noindent is the dimensionless kinetic energy for $p_x = p_y = 0$ with $a^2 = a_x^2 + a_y^2$.

\section{\label{sec:hydro}Description of One-Dimensional Beam Propagation for a Uniform Phase-Space Density}

It can be verified that there exist a class of exact solutions to the one-dimensional Vlasov equation (\ref{Onedimvlasov}) of the form \cite{Tzenov,TzenovBOOK}
\begin{equation}
F {\left( s, p_s; \tau \right)} = {\cal C} {\left\{ \Theta {\left[ p_s - p_s^{(-)} {\left( s; \tau \right)} \right]} - \Theta {\left[ p_s - p_s^{(+)} {\left( s; \tau \right)} \right]} \right\}}, \label{UniformDen}
\end{equation}
\noindent where $\Theta (z)$ is the well-known Heaviside step function. Expression (\ref{UniformDen}) implies that the distribution function has a constant phase-space density defined by the constant ${\cal C}$ within a simply connected region confined by the curves $p_s^{(-)} {\left( s; \tau \right)}$ and $p_s^{(+)} {\left( s; \tau \right)}$ and is zero outside. If $F {\left( s, p_s; \tau \right)}$ satisfies Eq. (\ref{UniformDen}) initially at $\tau = 0$, then the nonlinear Vlasov equation (\ref{Onedimvlasov}) assures that the phase-space density remains constant \cite{TzenovBOOK} at subsequent values of $\tau$ as the boundary curves $p_s^{(-)} {\left( s; \tau \right)}$ and $p_s^{(+)} {\left( s; \tau \right)}$ distort and evolve nonlinearly in response to the applied external and the self-generated fields. Clearly, the area number density
\begin{equation}
N_b = \int {\rm d} s {\rm d} p_s F {\left( s, p_s; \tau \right)}, \label{AreaNumDen}
\end{equation}
\noindent is preserved.

Next, we derive the evolution equations for the boundary curves $p_s^{(-)} {\left( s; \tau \right)}$ and $p_s^{(+)} {\left( s; \tau \right)}$. Substituting the explicit form of the solution (\ref{UniformDen}) into the Vlasov equation (\ref{Onedimvlasov}), we obtain
\begin{eqnarray}
- \delta {\left( p_s - p_s^{(-)} \right)} {\frac {\partial p_s^{(-)}} {\partial \tau}} + \delta {\left( p_s - p_s^{(+)} \right)} {\frac {\partial p_s^{(+)}} {\partial \tau}} \nonumber
\end{eqnarray}
\begin{eqnarray}
+ {\frac {p_s} {\gamma_c}} {\left[ - \delta {\left( p_s - p_s^{(-)} \right)} {\frac {\partial p_s^{(-)}} {\partial s}} + \delta {\left( p_s - p_s^{(+)} \right)} {\frac {\partial p_s^{(+)}} {\partial s}} \right]} \nonumber
\end{eqnarray}
\begin{eqnarray}
+ {\frac {\partial \gamma_c} {\partial s}} {\left[ - \delta {\left( p_s - p_s^{(-)} \right)} + \delta {\left( p_s - p_s^{(+)} \right)} \right]} \nonumber
\end{eqnarray}
\begin{equation}
= {\cal F} {\left[ - \delta {\left( p_s - p_s^{(-)} \right)} + \delta {\left( p_s - p_s^{(+)} \right)} \right]}. \label{Equunifpsden}
\end{equation}
\noindent Multiplying the above equation by $1$, $p_s$ and $p_s^2$ and integrating over $p_s$, we find the evolution equations for the boundary curves in the form
\begin{equation}
{\frac {\partial} {\partial \tau}} {\left( p_s^{(+)} - p_s^{(-)} \right)} + {\frac {\partial} {\partial s}} {\left( \gamma_c^{(+)} - \gamma_c^{(-)} \right)} = 0, \label{Boundary1}
\end{equation}
\begin{eqnarray}
{\frac {1} {2}} {\frac {\partial} {\partial \tau}} {\left( p_s^{(+) {\bf 2}} - p_s^{(-) {\bf 2}} \right)} + p_s^{(+)} {\frac {\partial \gamma_c^{(+)}} {\partial s}} - p_s^{(-)} {\frac {\partial \gamma_c^{(-)}} {\partial s}} \nonumber
\end{eqnarray}
\begin{equation}
= {\left( p_s^{(+)} - p_s^{(-)} \right)} {\cal F}, \label{Boundary2}
\end{equation}
\begin{eqnarray}
{\frac {1} {3}} {\frac {\partial} {\partial \tau}} {\left( p_s^{(+) {\bf 3}} - p_s^{(-) {\bf 3}} \right)} + p_s^{(+) {\bf 2}} {\frac {\partial \gamma_c^{(+)}} {\partial s}} - p_s^{(-) {\bf 2}} {\frac {\partial \gamma_c^{(-)}} {\partial s}} \nonumber
\end{eqnarray}
\begin{equation}
= {\left( p_s^{(+) {\bf 2}} - p_s^{(-) {\bf 2}} \right)} {\cal F}, \label{Boundary3}
\end{equation}
\noindent where in accordance with Eq. (\ref{Gammac}), we have defined
\begin{equation}
\gamma_c^{(\pm)} {\left( s; \tau \right)} = {\sqrt{1 + p_s^{(\pm)} {\left( s; \tau \right)}^2 + a^2 {\left( s; \tau \right)}}}. \label{Gammacpm}
\end{equation}
\noindent Let us now introduce the hydrodynamic variables $n$, $V$ and $\Gamma$ as
\begin{equation}
{\cal C} {\left( p_s^{(+)} - p_s^{(-)} \right)} = n \Gamma, \qquad {\cal C} {\left( \gamma_c^{(+)} - \gamma_c^{(-)} \right)} = n V \Gamma, \label{Hydrodynvar1}
\end{equation}
\begin{equation}
{\frac {\cal C} {2}} {\left( p_s^{(+) {\bf 2}} - p_s^{(-) {\bf 2}} \right)} = {\frac {\cal C} {2}} {\left( \gamma_c^{(+) {\bf 2}} - \gamma_c^{(-) {\bf 2}} \right)} = n V \Gamma^2. \label{Hydrodynvar2}
\end{equation}
\noindent Here $n {\left( s; \tau \right)}$ is the number density, $V {\left( s; \tau \right)}$ is the macroscopic flow velocity and $\Gamma {\left( s; \tau \right)}$ is yet unknown function to be specified later. From Eqs. (\ref{Hydrodynvar1}) and (\ref{Hydrodynvar2}), we readily obtain 
\begin{equation}
p_s^{(+)} = \Gamma {\left( V + {\frac {n} {2 {\cal C}}} \right)}, \qquad p_s^{(-)} = \Gamma {\left( V - {\frac {n} {2 {\cal C}}} \right)}, \label{Hydrodynvar3}
\end{equation}
\begin{equation}
\gamma_c^{(+)} = \Gamma {\left( 1 + {\frac {n V} {2 {\cal C}}} \right)}, \quad \gamma_c^{(-)} = \Gamma {\left( 1 - {\frac {n V} {2 {\cal C}}} \right)}. \label{Hydrodynvar4}
\end{equation}
\noindent Note from the above equations that the flow velocity $V$ and the function $\Gamma$ are given by
\begin{equation}
\Gamma V = {\frac {1} {2}} {\left( p_s^{(+)} + p_s^{(-)} \right)}, \qquad \Gamma = {\frac {1} {2}} {\left( \gamma_c^{(+)} + \gamma_c^{(-)} \right)}. \label{Hydrodynvar5}
\end{equation}
\noindent From the second of Eqs. (\ref{Hydrodynvar5}) with due account of Eqs. (\ref{Hydrodynvar3}) and (\ref{Hydrodynvar4}), we can express the function $\Gamma$ as
\begin{equation}
\Gamma = {\sqrt{\frac {1 + a^2} {{\left( 1 - V^2 \right)} {\left( 1 - 2 v_T^2 n^2 \right)}}}}, \label{Gammafunct}
\end{equation}
\noindent where
\begin{equation}
v_T^2 = {\frac {1} {8 {\cal C}^2}}, \label{Thermveloc}
\end{equation}
\noindent is the thermal speed squared.

To complete the macroscopic fluid description, we need to express the source terms entering the corresponding wave equations for the electromagnetic potentials as functions of $n$, $V$ and $\Gamma$. In the Lorentz gauge
\begin{equation}
{\frac {\partial a_s} {\partial s}} + {\frac {\partial \varphi} {\partial \tau}} = 0, \label{Lorentz}
\end{equation}
\noindent the scalar potential $\varphi$ and the vector potential ${\bf a}$ satisfy the wave equations
\begin{equation}
\square \varphi = {\frac {e^2} {\epsilon_0 m c^2}} \int {\rm d} p_s F {\left( s, p_s; \tau \right)}, \label{ScalarPot}
\end{equation}
\begin{eqnarray}
\square {\bf a}_{\perp} = {\frac {\mu_0 e^2} {m}} {\bf a}_{\perp} \int {\frac {{\rm d} p_s} {\gamma_c}} F {\left( s, p_s; \tau \right)} \nonumber
\end{eqnarray}
\begin{equation}
+ {\frac {\omega_c k_0} {c}} {\left( {\bf e}_x \cos k_0 s + {\bf e}_y \sin k_0 s \right)}, \label{VectorPotrans}
\end{equation}
\begin{equation}
\square a_s = {\frac {\mu_0 e^2} {m}} \int {\rm d} p_s {\frac {p_s} {\gamma_c}} F {\left( s, p_s; \tau \right)}, \label{VectorPotlong}
\end{equation}
\noindent where $\square = \partial_s^2 - \partial_{\tau}^2$ is the well-known d'Alembert operator and
\begin{equation}
\omega_c = {\frac {e B_0} {m}}, \label{Cyclotfr}
\end{equation}
\noindent is the electron cyclotron frequency associated with the amplitude of the wiggler field. The integral on the right-hand-side of Eq. (\ref{VectorPotrans}) can be expressed as
\begin{eqnarray}
\int {\frac {{\rm d} p_s} {\gamma_c}} F {\left( s, p_s; \tau \right)} = {\cal C} \ln {\left( {\frac {p_s^{(+)} + \gamma_c^{(+)}} {p_s^{(-)} + \gamma_c^{(-)}}} \right)} \nonumber
\end{eqnarray}
\begin{equation}
= {\cal C} \ln {\left( 1 + {\frac {n} {2 {\cal C}}} \right)} - {\cal C} \ln {\left( 1 - {\frac {n} {2 {\cal C}}} \right)}, \label{Calculate}
\end{equation}
\noindent while the integrals on the right-hand-side of Eqs. (\ref{ScalarPot}) and (\ref{VectorPotlong}) are standard and yield simply $n \Gamma$ and $n V \Gamma$, respectively.

Expressing Eqs. (\ref{Boundary1}) and (\ref{Boundary2}) for the moments in terms of the hydrodynamic variables, we have
\begin{equation}
{\frac {\partial} {\partial \tau}} {\left( n \Gamma \right)} + {\frac {\partial} {\partial s}} {\left( n \Gamma V \right)} = 0, \label{Continuity}
\end{equation}
\begin{equation}
{\frac {\partial} {\partial \tau}} {\left( V \Gamma \right)} + {\frac {\partial \Gamma} {\partial s}} = {\cal F}, \label{Euler}
\end{equation}
\noindent and from Eq. (\ref{Boundary3}) we obtain yet a third equation
\begin{eqnarray}
{\frac {\partial} {\partial \tau}} {\left( n \Gamma^3 V^2 + {\frac {n^3 \Gamma^3} {12 {\cal C}^2}} \right)} + \Gamma^2 V^2 {\frac {\partial} {\partial s}} {\left( n \Gamma V \right)} \nonumber
\end{eqnarray}
\begin{equation}
+ 2 n \Gamma^2 V {\frac {\partial \Gamma} {\partial s}} + {\frac {n^2 \Gamma^2} {4 {\cal C}^2}} {\frac {\partial} {\partial s}} {\left( n \Gamma V \right)} = 2 n \Gamma^2 V {\cal F}, \label{Energy}
\end{equation}
\noindent which is a direct consequence of the first two Eqs. (\ref{Continuity}) and (\ref{Euler}). The latter implies that the hierarchy of macroscopic fluid equations is closed and Eqs. (\ref{Continuity}) and (\ref{Euler}) comprise a complete hydrodynamic closure, fully equivalent to the one-dimensional Vlasov equation (\ref{Onedimvlasov}).

The macroscopic fluid equations (\ref{Continuity}) and (\ref{Euler}) must be supplemented with the equations for the self-fields. The wave equations for the scalar potential $\varphi$ and for the longitudinal component of the vector potential $a_s$ can be written in a straightforward manner to give
\begin{equation}
\square \varphi = {\frac {e^2} {\epsilon_0 m c^2}} n \Gamma, \label{Wavescalar}
\end{equation}
\begin{equation}
\square a_s = {\frac {\mu_0 e^2} {m}} n V \Gamma. \label{Wavevector}
\end{equation}
\noindent Note that the integral in Eq. (\ref{Calculate}) can be expressed as
\begin{equation}
\int {\frac {{\rm d} p_s} {\gamma_c}} F {\left( s, p_s; \tau \right)} = n {\left( 1 + {\frac {2} {3}} v_T^2 n^2 \right)} + O {\left( {\frac {1} {{\cal C}^4}} \right)}. \label{Calculate1}
\end{equation}
\noindent Standard procedure is to formally apply the thermodynamic limit in which ${\cal C}$ grows infinitely together with the volume occupied by the electron beam such that the number density $n$ remains finite. In this approximation the thermal speed $v_T$ becomes infinitely small as well. To second order, we can rewrite Eq. (\ref{VectorPotrans}) as
\begin{eqnarray}
\square {\bf a}_{\perp} = {\frac {\mu_0 e^2 n} {m}} {\left( 1 + {\frac {2} {3}} v_T^2 n^2 \right)} {\bf a}_{\perp} \nonumber
\end{eqnarray}
\begin{equation}
+ {\frac {\omega_c k_0} {c}} {\left( {\bf e}_x \cos k_0 s + {\bf e}_y \sin k_0 s \right)}. \label{VectorPoten}
\end{equation}
\noindent In order to eliminate the explicit dependence of the right-hand-side of Eq. (\ref{VectorPoten}) on the longitudinal coordinate $s$, we introduce the helical field variables according to
\begin{equation}
{\cal A}_x = a_x \cos k_0 s + a_y \sin k_0 s, \label{HelicalVarx}
\end{equation}
\begin{equation}
{\cal A}_y = - a_x \sin k_0 s + a_y \cos k_0 s. \label{HelicalVary}
\end{equation}
\noindent Note that $a^2 = a_x^2 + a_y^2 = {\cal A}_x^2 + {\cal A}_y^2$. Finally, we collect the macroscopic fluid equations (\ref{Continuity}) and (\ref{Euler}) together with the equations for the electromagnetic fields to write the basic system
\begin{equation}
{\frac {\partial} {\partial \tau}} {\left( n \Gamma \right)} + {\frac {\partial} {\partial s}} {\left( n \Gamma V \right)} = 0, \label{Continuityf}
\end{equation}
\begin{equation}
{\frac {\partial} {\partial \tau}} {\left( V \Gamma \right)} + {\frac {\partial \Gamma} {\partial s}} = {\cal F}, \label{Eulerf}
\end{equation}
\begin{equation}
\square \varphi = {\frac {e^2} {\epsilon_0 m c^2}} n \Gamma, \label{Poissonf}
\end{equation}
\begin{equation}
\square {\cal A}_x - 2 k_0 {\frac {\partial {\cal A}_y} {\partial s}} - k_0^2 {\cal A}_x = {\frac {\mu_0 e^2 n {\cal A}_x} {m}} {\left( 1 + {\frac {2} {3}} v_T^2 n^2 \right)} + {\frac {\omega_c k_0} {c}}, \label{VectorPotenx}
\end{equation}
\begin{equation}
\square {\cal A}_y + 2 k_0 {\frac {\partial {\cal A}_x} {\partial s}} - k_0^2 {\cal A}_y = {\frac {\mu_0 e^2 n {\cal A}_y} {m}} {\left( 1 + {\frac {2} {3}} v_T^2 n^2 \right)}. \label{VectorPoteny}
\end{equation}
\begin{equation}
\square a_s = {\frac {\mu_0 e^2} {m}} n V \Gamma. \label{Wavevectorf}
\end{equation}
\noindent of equations, which will be the starting point for the subsequent analysis.

Worthwhile to mention is that, instead of the Lorentz gauge adopted in the present paper, it is possible to use the Coulomb gauge, where $\partial_s a_s = 0$. In this case it is necessary to take Eq. (\ref{Poissonf}) in the form
\begin{equation}
\nabla^2 \varphi = \partial_s^2 \varphi = {\frac {e^2} {\epsilon_0 m c^2}} n \Gamma, \label{PoissonCoul}
\end{equation}
while Eq. (\ref{Wavevectorf}) transforms as
\begin{equation}
\square a_s - \partial_{\tau} \partial_s \varphi = - \partial_{\tau} {\cal F} = {\frac {\mu_0 e^2} {m}} n V \Gamma. \label{WavevectorCoul}
\end{equation}

It is important to note that the exact hydrodynamic model derived here is invariant under Lorentz transformation. Detailed proof of this assertion is given in Appendix \ref{sec:appendix}.

\section{\label{sec:linear}Linear Stability of the Relativistic Hydrodynamic Model}

Let us consider in what follows the simplest case of a cold electron beam, that is the limiting case, where $n v_T \rightarrow 0$. First of all, we note that the hydrodynamic equations (\ref{Continuityf}) and (\ref{Eulerf}) admit a stationary solution of the form $n = n_0 = {\rm const}$ and $V = v_0 = {\rm const}$. In addition, the stationary transverse components of the vector potential are given by the expressions
\begin{equation}
{\cal A}_{x0} = - {\frac {\omega_c k_0 c} {\omega_p^2 + k_0^2 c^2}}, \qquad \quad {\cal A}_{y0} = 0, \label{StatVectorTrans}
\end{equation}
where $\omega_p$ is the plasma frequency expressed as
\begin{equation}
\omega_p^2 = {\frac {e^2 n_0} {\epsilon_0 m}}. \label{PlasmaFreq}
\end{equation}
Note also that in obtaining the stationary solutions above, the explicit assumption that the beam density $n_0$ is small has been made. Therefore, the longitudinal space-charge effects can be neglected ($\varphi_0$, $a_{s0}$ and ${\cal F}_0$ are all zero). Finally,
\begin{equation}
\Gamma_0 = \gamma_0 R_0, \qquad \gamma_0 = {\frac {1} {\sqrt{1 - v_0^2}}}, \qquad R_0 = {\sqrt{1 + {\cal A}_{x0}^2}}. \label{StatGamma}
\end{equation}

Instead of the full system of equations (\ref{Continuityf}) - (\ref{Wavevectorf}), we shall use
\begin{equation}
{\frac {\partial} {\partial \tau}} {\left( n \Gamma \right)} + {\frac {\partial} {\partial s}} {\left( n \Gamma V \right)} = 0, \label{ContinuityStat}
\end{equation}
\begin{equation}
{\frac {\partial^2} {\partial \tau \partial s}} {\left( V \Gamma \right)} + {\frac {\partial^2 \Gamma} { \partial s^2}} = {\frac {e^2} {\epsilon_0 m c^2}} n \Gamma, \label{EulerStat}
\end{equation}
coupled with the equations for the transverse vector potential (\ref{VectorPotenx}) and (\ref{VectorPoteny}). Equation (\ref{EulerStat}) has been obtained by differentiating Eq. (\ref{Eulerf}) with respect to the longitudinal coordinate $s$ and making use of the Lorentz gauge condition (\ref{Lorentz}) and the equation for the scalar potential (\ref{Poissonf}).

Following the standard procedure, we take
\begin{equation}
n = n_0 + \epsilon n_1 + \dots, \qquad V = v_0 + \epsilon v_1 + \dots, \label{LinearStat1}
\end{equation}
\begin{equation}
{\cal A}_x = {\cal A}_{x0} + \epsilon {\cal A}_{x1} + \dots, \qquad {\cal A}_y = \epsilon {\cal A}_{y1} + \dots, \label{LinearStat2}
\end{equation}
and expand the quantities $\Gamma$, $\Gamma V$, $n \Gamma$ and $n \Gamma V$ to first order in the formal expansion parameter $\epsilon$. Substituting all of the above into Eqs. (\ref{ContinuityStat}), (\ref{EulerStat}), (\ref{VectorPotenx}) and (\ref{VectorPoteny}), and retaining linear terms, we obtain
\begin{equation}
{\left( \partial_{\tau} + v_0 \partial_s \right)} n_1 + n_0 \gamma_0^2 {\left( v_0 \partial_{\tau} + \partial_s \right)} v_1 \nonumber
\end{equation}
\begin{equation}
+ {\frac {n_0 {\cal A}_{x0}} {R_0^2}} {\left( \partial_{\tau} + v_0 \partial_s \right)} {\cal A}_{x1} = 0, \label{LinStatCont}
\end{equation}
\begin{equation}
\gamma_0^2 \partial_s {\left( \partial_{\tau} + v_0 \partial_s \right)} v_1 + {\frac {{\cal A}_{x0}} {R_0^2}} \partial_s {\left( v_0 \partial_{\tau} + \partial_s \right)} {\cal A}_{x1} \nonumber
\end{equation}
\begin{equation}
= {\frac {\omega_p^2} {c^2}} {\left( {\frac {{\cal A}_{x0}} {R_0^2}} {\cal A}_{x1} + v_0 \gamma_0^2 v_1 + {\frac {n_1} {n_0}} \right)}, \label{LinStatEuler}
\end{equation}
\begin{equation}
\square {\cal A}_{x1} - 2 k_0 \partial_s {\cal A}_{y1} - k_1^2 {\cal A}_{x1} = {\frac {\omega_p^2 {\cal A}_{x0}} {c^2 n_0}} n_1, \label{LinStatVecPotx}
\end{equation}
\begin{equation}
\square {\cal A}_{y1} + 2 k_0 \partial_s {\cal A}_{x1} - k_1^2 {\cal A}_y = 0, \label{LinStatVecPoty}
\end{equation}
where
\begin{equation}
k_1^2 = k_0^2 + {\frac {\omega_p^2} {c^2}}. \label{ShiftK0}
\end{equation}
It is convenient to perform a Lorentz transformation by introducing a new time and longitudinal coordinate variables according to the expressions
\begin{equation}
\theta = \gamma_0 {\left( \tau - v_0 s \right)}, \qquad z = \gamma_0 {\left( s - v_0 \tau \right)}. \label{LorentzTrans}
\end{equation}
Derivatives transform according to
\begin{equation}
\partial_{\tau} = \gamma_0 {\left( \partial_{\theta} - v_0 \partial_z \right)}, \qquad \partial_s = \gamma_0 {\left( - v_0 \partial_{\theta} + \partial_z \right)}. \label{LorentzDeriv}
\end{equation}
Differential operators entering Eqs. (\ref{LinStatCont}) and (\ref{LinStatEuler}) simplify considerably
\begin{equation}
\partial_{\tau} + v_0 \partial_s = {\frac {1} {\gamma_0}} \partial_{\theta}, \qquad v_0 \partial_{\tau} + \partial_s = {\frac {1} {\gamma_0}} \partial_z, \label{LorentzSimpl}
\end{equation}
while the d'Alembert operator $\square$ is invariant in the inertial frame moving in the longitudinal direction at speed $v_0$. Thus, in the new Lorentz coordinate system the linearized equations simplify considerably
\begin{equation}
\partial_{\theta} n_1 + n_0 \gamma_0^2 \partial_z v_1 + {\frac {n_0 {\cal A}_{x0}} {R_0^2}} \partial_{\theta} {\cal A}_{x1} = 0, \label{LinLorCont}
\end{equation}
\begin{equation}
\gamma_0^2 \partial_{\theta} {\left( - v_0 \partial_{\theta} + \partial_z \right)} v_1 + {\frac {{\cal A}_{x0}} {R_0^2}} \partial_z {\left( - v_0 \partial_{\theta} + \partial_z \right)} {\cal A}_{x1} \nonumber
\end{equation}
\begin{equation}
= {\frac {\omega_p^2} {c^2}} {\left( {\frac {{\cal A}_{x0}} {R_0^2}} {\cal A}_{x1} + v_0 \gamma_0^2 v_1 + {\frac {n_1} {n_0}} \right)}, \label{LinLorEuler}
\end{equation}
\begin{equation}
{\left( \square - k_1^2 \right)} {\cal A}_{x1} - 2 k_0 \gamma_0 {\left( - v_0 \partial_{\theta} + \partial_z \right)} {\cal A}_{y1} = {\frac {\omega_p^2 {\cal A}_{x0}} {c^2 n_0}} n_1, \label{LinLorVecPotx}
\end{equation}
\begin{equation}
{\left( \square - k_1^2 \right)} {\cal A}_{y1} + 2 k_0 \gamma_0 {\left( - v_0 \partial_{\theta} + \partial_z \right)} {\cal A}_{x1} = 0. \label{LinLorVecPoty}
\end{equation}
Manipulating the first two Eqs. (\ref{LinLorCont}) and (\ref{LinLorEuler}), we can eliminate the linear velocity $v_1$ and obtain a single equation relating the linear density $n_1$ and the horizontal component of the vector potential. We have
\begin{equation}
{\left( - v_0 \partial_{\theta} + \partial_z \right)} {\left[ {\frac {{\cal A}_{x0}} {R_0^2}} {\left( \square - {\frac {\omega_p^2} {c^2}} \right)} {\cal A}_{x1} - {\left( \partial_{\theta}^2 + {\frac {\omega_p^2} {c^2}} \right)} {\frac {n_1} {n_0}} \right]} = 0. \label{Eliminate}
\end{equation}
It is now a simple matter to obtain a single equation for the horizontal component of the vector potential. The result is
\begin{equation}
{\left( - v_0 \partial_{\theta} + \partial_z \right)} {\widehat{\aleph}} {\cal A}_{x1} = 0, \label{DispersionEq}
\end{equation}
where the operator ${\widehat{\aleph}}$ is given by
\begin{equation}
{\widehat{\aleph}} = {\left( \partial_{\theta}^2 + {\frac {\omega_p^2} {c^2}} \right)} {\left( \square - k_1^2 \right)}^2 + 4 k_0^2 \gamma_0^2 {\left( \partial_{\theta}^2 + {\frac {\omega_p^2} {c^2}} \right)} \nonumber
\end{equation}
\begin{equation}
\times {\left( - v_0 \partial_{\theta} + \partial_z \right)}^2 - {\frac {\omega_p^2 {\cal A}_{x0}^2} {c^2 R_0^2}} {\left( \square - k_1^2 \right)} {\left( \square - {\frac {\omega_p^2} {c^2}} \right)}. \label{Operator}
\end{equation}

Note that, the left-hand-side of Eq. (\ref{DispersionEq}) represents a product of two operators. In the laboratory frame the action of the first operator ${\left( - v_0 \partial_{\theta} + \partial_z \right)}$ simply implies that the linear solution ${\cal A}_{x1}$ to our fluid dynamic model does not depend on the longitudinal variable $s$ and therefore is an arbitrary function of time to this end. This observation is consistent with Eqs. (\ref{LinStatCont}) and (\ref{LinStatEuler}), provided the condition
\begin{equation}
{\frac {{\cal A}_{x0}} {R_0^2}} {\cal A}_{x1} + v_0 \gamma_0^2 v_1 + {\frac {n_1} {n_0}} = 0, \label{Condition}
\end{equation}
holds for the time dependent first order quantities. Note also that, if the velocity $v_1 {\left( \tau \right)}$ is chosen arbitrarily (but dependent on time $\tau$ only), then $n_1 {\left( \tau \right)}$ and ${\cal A}_{x1} {\left( \tau \right)}$ can be determined uniquely from Eqs. (\ref{LinStatVecPotx}) and (\ref{Condition}). Namely, expressing $n_1$ in terms of ${\cal A}_{x1}$ and $v_1$ from Eq. (\ref{Condition}) and substituting the result into Eq. (\ref{LinStatVecPotx}), we obtain
\begin{equation}
\partial_{\tau}^2 {\cal A}_{x1} + \omega_h^2 {\cal A}_{x1} = {\frac {\omega_p^2} {c^2}} {\cal A}_{x0} v_0 \gamma_0^2 v_1, \label{PlasmonEq}
\end{equation}
where
\begin{equation}
\omega_h^2 = k_0^2 + {\frac {\omega_p^2} {c^2 R_0^2}}, \label{CharFrequency}
\end{equation}
is the characteristic frequency of harmonic oscillations.

In the next Section, we analyze in detail the Fourier spectrum of the operator ${\widehat{\aleph}}$.

\section{\label{sec:simulation}Numerical Results for the Small-Signal Gain}

Let us rewrite the operator defined in Eq. (\ref{Operator}) in the laboratory frame as follows
\begin{equation}
{\widehat{\aleph}} = {\left[ \gamma_0^2 {\left( \partial_{\tau} + v_0 \partial_s \right)}^2 + {\frac {\omega_p^2} {c^2}} \right]} {\left[ {\left( \square - k_1^2 \right)}^2 + 4 k_0^2 \partial_s^2 \right]} \nonumber
\end{equation}
\begin{equation}
- {\frac {\omega_p^2 {\cal A}_{x0}^2} {c^2 R_0^2}} {\left( \square - k_1^2 \right)} {\left( \square - {\frac {\omega_p^2} {c^2}} \right)}. \label{OperatorLab}
\end{equation}
Next, we seek a solution to the equation ${\widehat{\aleph}} {\cal A}_{x1} = 0$ in the standard form
\begin{equation}
{\cal A}_{x1} = \sum_n \int \limits_{- \infty}^{\infty} {\rm d} k {\cal A} {\left( k \right)} \exp {\left[ i k s - i \omega_n {\left( k \right)} \tau \right]}, \label{LinearSol}
\end{equation}
where $\omega_n {\left( k \right)}$ are all possible solutions of the {\it dispersion equation}
\begin{equation}
{\left[ {\frac {\omega_p^2} {c^2}} - \gamma_0^2 {\left( k v_0 - \omega \right)}^2 \right]} \nonumber
\end{equation}
\begin{equation}
\times {\left[ \omega^2 - {\left( k + k_0 \right)}^2 - {\frac {\omega_p^2} {c^2}} \right]} {\left[ \omega^2 - {\left( k - k_0 \right)}^2 - {\frac {\omega_p^2} {c^2}} \right]} \nonumber
\end{equation}
\begin{equation}
= {\frac {\omega_p^2 {\cal A}_{x0}^2} {c^2 R_0^2}} {\left( \omega^2 - k^2 - k_1^2 \right)} {\left( \omega^2 - k^2 - {\frac {\omega_p^2} {c^2}} \right)}. \label{DisperEquation}
\end{equation}

\begin{figure}
\begin{center}
\includegraphics[width=8.0cm]{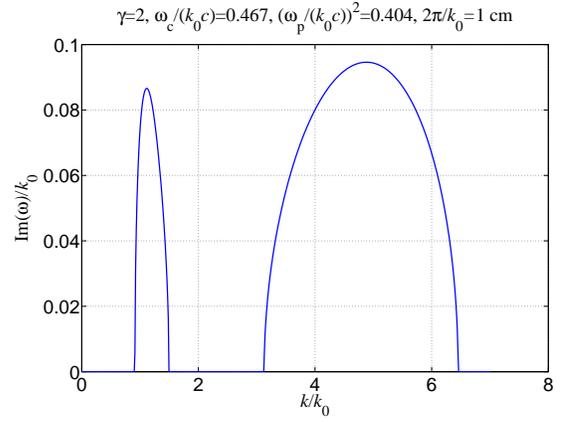}
\caption{\label{fig:epsart1} Free electron laser instability growth rate as a function of the normalized radiation wave number.}
\end{center}
\end{figure}

\begin{figure}
\begin{center}
\includegraphics[width=8.0cm]{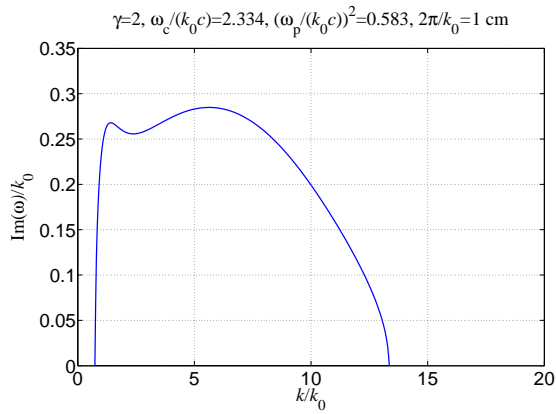}
\caption{\label{fig:epsart2} The same as in Fig. \ref{fig:epsart1} but at higher magnetic field and beam density values.}
\end{center}
\end{figure}

\begin{figure}
\begin{center}
\includegraphics[width=8.0cm]{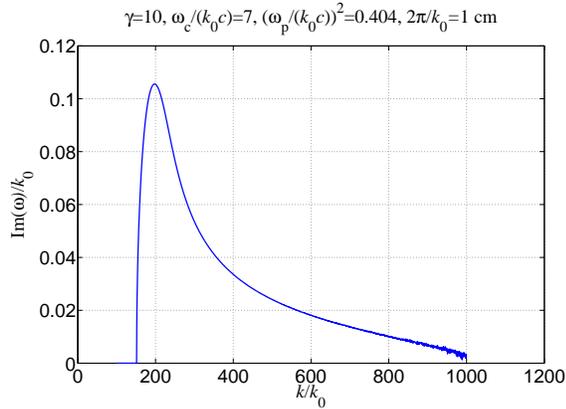}
\caption{\label{fig:epsart3} The same as Fig. \ref{fig:epsart1} but at higher beam energy and magnetic field strength.}
\end{center}
\end{figure}

\begin{figure}
\begin{center}
\includegraphics[width=8.0cm]{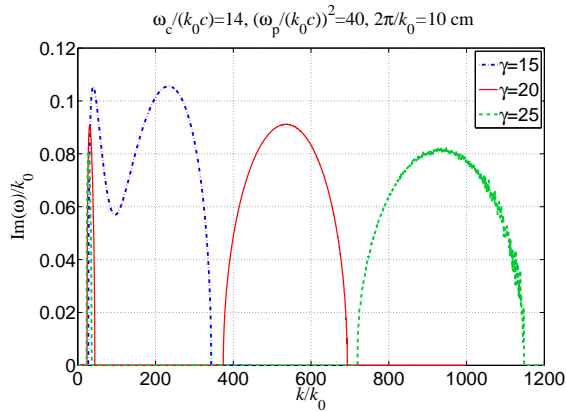}
\caption{\label{fig:epsart4} Dependence of the instability growth rate on beam energy for fixed magnetic field, beam density and undulator period values.}
\end{center}
\end{figure}

\begin{figure}
\begin{center}
\includegraphics[width=8.0cm]{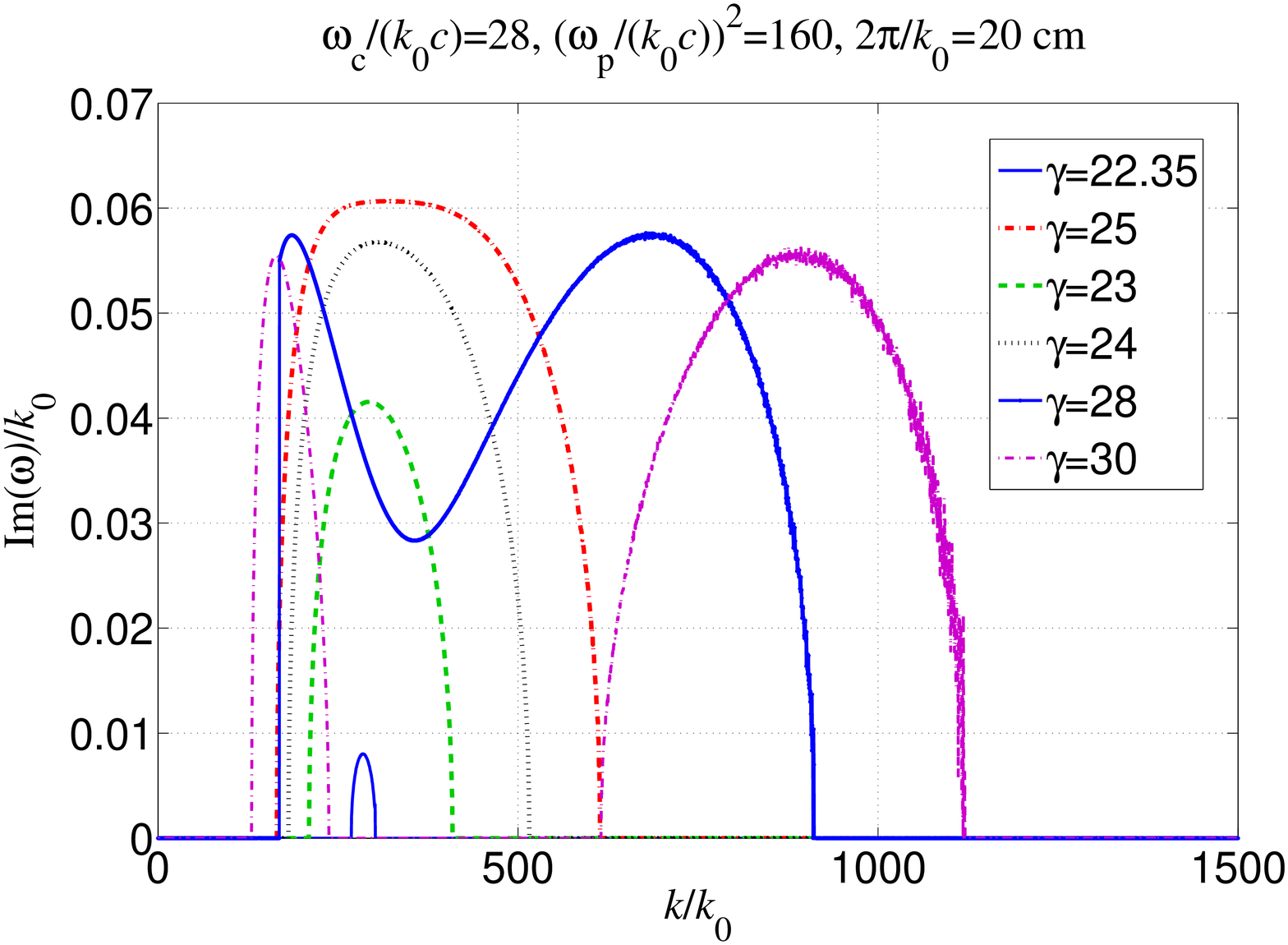}
\caption{\label{fig:epsart5} The same as in Fig. \ref{fig:epsart4} but the undulator period length has been increased by a factor of 2.}
\end{center}
\end{figure}

In order to assess the predictions of the hydrodynamic model on a qualitative and quantitative level, Eq. (\ref{DisperEquation}) has been solved numerically and the results are presented in Figures \ref{fig:epsart1} -- \ref{fig:epsart5}. Note, that in the notation adopted here the frequency $[\omega] = m^{-1}$ is, in fact, the radiation wave number.

At low $\gamma_0$ values (Figures \ref{fig:epsart1} and \ref{fig:epsart2}) the observed behavior of the free electron laser instability growth rate is qualitatively similar to that reported earlier \cite{DavidsonUhm}. Note, however, that a hydrodynamic approach is adopted here, whereas a kinetic description has been used by Davidson and Uhm \cite{DavidsonUhm}. In addition there is a gamma factor difference in the definitions of $\omega_p$ and $\omega_c$ in the paper by Davidson and Uhm \cite{DavidsonUhm} and here.

At higher $\gamma_0$ and undulator magnetic field strength values (see Figure \ref{fig:epsart3}) the instability bandwidth increases and the location of its peak shifts towards shorter radiation wavelengths.

Figures \ref{fig:epsart4} and \ref{fig:epsart5} show the effect of beam energy on the instability at two different undulator period lengths. As can be seen at lower beam energies a single instability band exists, whereas at higher beam energies two separate bands are generated. In addition (see Figure \ref{fig:epsart5}), there is a threshold beam energy below which no instability exists. Increasing beam energy above its threshold value an increase of both the bandwidth (range of $k$-values) and the peak value of $\omega$ as a function of $k$ can be observed.

\section{\label{sec:remarks}Concluding Remarks}

Starting from first principle an exact relativistic hydrodynamic closure of equations describing the dynamics of an electron beam propagating axially in a static magnetic undulator field has been obtained. It has been shown that the hydrodynamic equations are fully equivalent to the Vlasov-Maxwell system for a special type of solutions of the Vlasov equation. Moreover, as expected, the warm (in general) fluid dynamic equations derived in the present paper are invariant under Lorentz transformation.

Another interesting feature of our hydrodynamic picture is the underlying pressure law. The latter together with the ponderomotive potential (usually proportional to the transverse vector potential squared) represents an effective enthalpy of the system [see Eq. (\ref{Gammafunct})]. Noteworthy to mention is also that in the non relativistic limit our system of hydrodynamic equations reduces to the well-known picture with triple adiabatic pressure law \cite{Tzenov,TzenovBOOK}.

As a direct application of the theory developed here, the linear stability of the exact hydrodynamic closure has been studied. It has been shown that all basic features of the small signal gain can be derived from the fluid dynamic description developed in the present paper. Satisfactory agreement with previously reported results has been also found.

A possible extension of the approach initiated here could be a numerical modeling of the hydrodynamic equations, as well as analysis of nonlinear effects and possible formation of solitary wave patterns and coherent structures. All of the above we plan to perform in the near future.

\appendix

\section{\label{sec:appendix}Lorentz Invariance of the Hydrodynamic Model}

Let us first represent the quantity $\Gamma$ defined by Eq. (\ref{Gammafunct}) as
\begin{equation}
\Gamma = \gamma \Sigma, \label{GammaRepres}
\end{equation}
where
\begin{equation}
\gamma = {\frac {1} {\sqrt{1 - V^2}}}, \qquad \quad \Sigma = {\sqrt{\frac {1 + a^2} {1 - 2 v_T^2 n^2}}}. \label{GammaSigma}
\end{equation}
In order to prove the relativistic invariance of our basic system of equations (\ref{Continuityf}) - (\ref{Wavevectorf}), we recall the Lorentz velocity addition law
\begin{equation}
V = {\frac {V^{\prime} + v_0} {1 + v_0 V^{\prime}}}, \qquad \quad \gamma = \gamma_0 \gamma^{\prime} {\left( 1 + v_0 V^{\prime} \right)}, \label{LorentzVel}
\end{equation}
following from the Lorentz transformation (\ref{LorentzTrans}). Here, with prime we denote the value of the corresponding quantity in a coordinate system moving in the longitudinal direction $s$ with velocity $v_0$. In addition, the scalar potential $\varphi$ and the longitudinal component of the vector potential $a_s$ transform according to the expressions
\begin{equation}
\varphi^{\prime} = \gamma_0 {\left( \varphi - v_0 a_s \right)} , \qquad \quad a_s^{\prime} =  \gamma_0 {\left( a_s - v_0 \varphi \right)}, \label{ElectroMag}
\end{equation}

Using expressions (\ref{LorentzDeriv}) for the corresponding derivatives, we obtain
\begin{equation}
\gamma_0^2 {\left( \partial_{\theta} - v_0 \partial_z  \right)} {\left[ n \Gamma^{\prime} {\left( 1 + v_0 V^{\prime} \right)} \right]} \nonumber
\end{equation}
\begin{equation}
+ \gamma_0^2 {\left( - v_0 \partial_{\theta} + \partial_z  \right)} {\left[ n \Gamma^{\prime} {\left( V^{\prime} + v_0 \right)} \right]} = 0. \label{Express}
\end{equation}
Simple rearrangement of terms in the above equation leads to Eq. (\ref{Continuityf}) in the new coordinate system. In a similar way it can be verified that the left-hand-side of the momentum balance equation (\ref{Eulerf}) does not change in the moving system. Since the longitudinal force ${\cal F}$ is an obvious Lorentz invariant due to the transformation law of the electromagnetic potentials (\ref{ElectroMag}), Eq. (\ref{Eulerf}) also remains unchanged in the new coordinate system.


\nocite{*}
\bibliography{aipsamp}

\begin{thebibliography}{99}
\bibitem{Saldin} E. Saldin, A. Schneidmiller and M. Yurkov, "{\it The Physics of Free Electron Lasers}", Springer (2000).
\bibitem{Freund} H. Freund and T. Antonsen, "{\it Principles of Free Electron Lasers}", Chapman and Hall (1996).
\bibitem{Clarke} J. Clarke, "{\it The Science and Technology of Undulators and Wigglers}", Oxford University Press (2004).
\bibitem{Madey} J. M. J. Madey, Journal of Applied Physics, {\bf 42}, 1906 (1971).
\bibitem{Colson} W. Colson, Physics Letters {\bf A}, {\bf 59}, 187 (1976).
\bibitem{Becker} W. Becker and H. Mitter, Zeitschrift fur Physik {\bf B}, {\bf 35}, 399 (1979).
\bibitem{Bambini} A. Bambini, A. Renieri and S. Stenholm,  Physical Review {\bf A}, {\bf 19}, 2013 (1979).
\bibitem{Kroll} N. M. Kroll and W. A. McMullin, Physical Review {\bf A}, {\bf 17}, 300, (1978).
\bibitem{Kwan} T. Kwan, J. M. Dawson, and A. T. Lin, Physics of Fluids, {\bf 20}, 581 (1977).
\bibitem{Ride} W. Colson and S. K. Ride, Physics Letters {\bf A}, {\bf 76}, 379 (1980).
\bibitem{Hopf} H. Al-Abawi, F. A. Hopf, G. T. Moore and M. 0. Scully,  Optics Communications, {\bf 30}, 235 (1979).
\bibitem{Davidson} R. C. Davidson, Physics of Fluids, {\bf 29}, 2689 (1986).
\bibitem{DavidsonUhm} R. C. Davidson and H. S. Uhm, Physics of Fluids, {\bf 23}, 2076 (1980).
\bibitem{Piovella} N. Piovella, P. Chaix, G. Shvets and D. A. Jaroszynski, Physical Review {\bf E}, {\bf 52}, 5470 (1995).
\bibitem{Bonifacio} R. Bonifacio, B. W. J. McNeil and P. Pierini, Physical Review {\bf A}, {\bf 40}, 4467 (1989).
\bibitem{Bonifacio1} R. Bonifacio, F. Casagrande, G. Gerchioni, L. de Salvo Souza, P. Pierini and N. Piovella, Nuovo Cimento, {\bf 13}, 1 (1990).
\bibitem{MarTze} K. B. Marinov and S. I. Tzenov, Physics of Plasmas {\bf 18}, 032305 (2011).
\bibitem{DavidsonSen} R. C. Davidson, G. L. Johnston and A. Sen, Physical Review {\bf A}, {\bf 34}, 392 (1986).
\bibitem{Sen} A. Sen and G. L. Johnston, Physical Review Letters, {\bf 70}, 786 (1993).
\bibitem{Tzenov} R. C. Davidson, H. Qin, S. I. Tzenov, and E. A. Startsev, Physical Review ST Accel. Beams {\bf 5}, 084402 (2002).
\bibitem{TzenovBOOK} S. I. Tzenov, ``{\it Contemporary Accelerator Physics}'', World Scientific (2004).




\end{thebibliography}

\end{document}